\documentclass[%
reprint,
 amsmath,amssymb,
 aps,
]{revtex4-2}

\usepackage{graphicx}
\usepackage{dcolumn}
\usepackage{bm}

\begin{document}

\preprint{APS/123-QED}

\title{Superselective adsorption of multivalent polymeric particles}

\author{Elena Patyukova}
 \email{patyukova@gmail.com}
 \altaffiliation{Chemistry Department, Durham University, Durham, DH1 3LE, UK}


\begin{abstract}
Multivalency is a common biological mechanism of formation of strong reversible and selective bonds by grouping weak bonds. Polymers often act as a scaffold to which multiple binding groups are attached. Here I present an analytical theory allowing to calculate avidity and selectivity for multivalent polymeric particles using tools from the theory of associating fluids. I explicitly take into account conformational degrees of freedom of a polymeric scaffold and discuss how they affect superselectivity. I also consider linear polymeric particles with two types of ligands and show that superselectivity does not depend on the sequence of ligands along the backbone for a Gaussian chain with short linkers. 
\end{abstract}

\keywords{superselective binding, multivalency, targeting, adsoption, associating fluids}
\maketitle


\section{Introduction}

The ability to switch between different states in a controllable manner and show high degree of binding selectivity are essential components of intelligent behavior of biological systems. Understanding the mechanisms of these phenomena from viewpoint of statistical physics is a rapidly developing area of science. Two important concepts introduced recently in this context are multivalency\cite{Mammen1998PolyvalentInhibitors, Fasting2012MultivalencyPrinciple} and superselectivity\cite{Martinez-Veracoechea2011DesigningBinding}. 

A multivalent particle is a particle which has multiple bonding sites ("ligands") which can bind selectively to another carrier decorated with complimentary bonding sites ("receptors"). Initially multivalent interactions were defined as a mechanism of translating individually low affinity interaction in to very strong and specific connections\cite{Mammen1998PolyvalentInhibitors}. 
Quantitatively this effect was characterised in the literature by the term "avidity", which describes a cumulative association constant for interaction between multivalent particles\cite{Krishnamurthy2006MultivalencyDesign}, and is usually several orders of magnitude larger than association constant characterising individual interaction between one ligand and one receptor. It was explained that the main factor leading to increase in avidity is a stronger than linear growth of a number of possible arrangements between ligands and receptors when their numbers increase\cite{Kitov2003OnModel}.

Later it was pointed out that enhanced strength of interaction on its own does not lead to recognition and the sharpness of transition between bonded and unbonded state in an ensemble of multivalent particles is important\cite{Martinez-Veracoechea2011DesigningBinding}. Martinez-Veracoechea and Frenkel considered the specific case of adsorption of multivalent particles on a plane with some density of receptors. They postulated that particles can recognise a plane with a given density of receptors if the adsorption transition is sufficiently sharp\cite{Martinez-Veracoechea2011DesigningBinding}. In order to characterise the sharpness of adsorption transition they introduced "selectivity", defined as 
\begin{equation}
    \alpha=\frac{\partial \ln \theta}{\partial \ln n_R}
\end{equation}
where $\theta$ is the fraction of bonded surface sites and $n_R$ is the number of receptors per site (which is proportional to the density of receptors on the surface).
Adsorption is called superselective if max$\left(\alpha\right)>1$. Martinez-Veracoechea and Frenkel showed using both theoretical and computational models that sharpness of the transition and max$\left(\alpha\right)$ indeed increase as the valency of particles grows. Here again the main source of sharpness of the transition is a non-linear increase of the number of possible arrangements between ligands and receptors.

A large class of multivalent particles developed for different applications are polymeric in nature \cite{VanDongen2014MultivalentConjugation}. At this point an interesting question arises whether superselectivity will be present in multivalent system in which ligands are non-equivalent as on a linear polymer. Non-equivalence of ligands is based in the fact that pinning a polymer to a surface has entropic cost which is dependent on the location of a segment along the chain. In this situation the theoretical model of Martinez-Veracoechea and Frenkel can not be directly applied to predict selectivity, because it assumes equal entropic costs of bond formation for all ligands of the multivalent particle. Here I develop an analytical approach allowing to relax this assumption. I apply developed approach to consider a superselective adsorption of a single polymer chain and study how selectivity depends on valency of polymer and length of linkers between stickers. I also look at how modification of some stickers to stronger or weaker ones affects selectivity curve. 

The paper is organised as follows: (1) firstly I show how selectivity formula for multivalent particles with no internal degrees of freedom previously derived by Martinez-Veracoechea and Frenkel\cite{Martinez-Veracoechea2011DesigningBinding} can be obtained in a new way; (2) then it is showed how to calculate additional entropy cost of bonds formation associated with deformation of a polymer chain; (3) finally I join these parts together to calculated selectivity for adsorption of a polymer chain with entropy cost of bonds formation taken into account. (4) I consider chains decorated with two types of stickers.
At the end I analyse obtained expressions and make conclusions.

\section{Adsorption isotherm for multivalent particles}

\begin{figure}
\includegraphics[width=0.7\columnwidth]{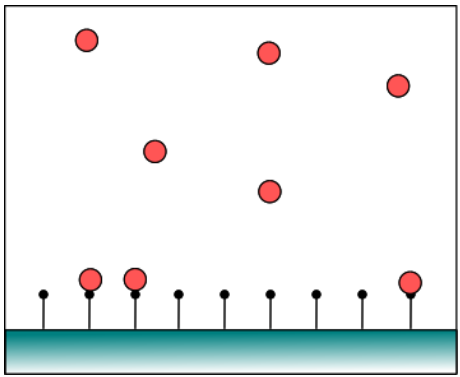}
\caption{Visual description of Langmuir adsorption model}
\label{langmuir}
\end{figure}

First of all let us remind ourselves that adsorption of univalent particles is described by Langmuir model (see Figure \ref{langmuir}). The Langmuir model implies the following assumptions: (1) particles do not interact with each other; (2) the surface is flat; (3) all sites are equivalent.
Let us notice that Langmuir isotherm can be obtained using association model approach common for description of associating fluids \cite{Veytsman1990a,Patyukova2018HydrogenExperiment}. In the association model approach formation of specific bonds, such as hydrogen bonds, is considered as a reversible chemical reaction. Then it is assumed that configurational part of partition function of the system can be factorized into a part describing non-specific interactions and a part describing contribution of specific interactions (reversible bonds). This factorization of the partition function has no rigorous explanation, however, it is usually said that it can be done because of large difference in timescales characterizing non-specific interactions and specific reversible bonds.  For Langmuir adsorption problem this means that we can write:
\begin{equation}
    \frac{F}{kT}=N\ln \frac{N\Lambda^3}{Ve}-\ln Z_{ad}
    \label{free_langmuir}
\end{equation}
Here the first term is the free energy of an ideal gas as far in Langmuir model particles do not interact with each other and the second term is contribution to free energy due to formation of bonds between particles and the surface. $N$ is the number of all particles in the system. Let us additionally denote as $N_s$ the number of surface bonding sites; $m$ the number of adsorbed particles; $f$ the free energy of formation of a bond between the particle and the surface which has both enthalpic contribution (which should be negative) and probably entropic contribution, i.e. accounting for loss of orientational entropy upon formation of a bond; $V$ the total volume of the system and $v$ the volume available to the the particle in the bonded state. We assume that all particles are univalent. Then the part of the partition function associated with formation of bonds $Z_{ad}$ can be calculated as:
\begin{equation}
    Z_{ad}=e^{-mf}\left(\frac{v}{V}\right)^m\frac{N!N_s!}{\left(N-m\right)!\left(N_s-m\right)!m!}
    \label{z_langmuir}
\end{equation}
The first factor describes the energy of formation of $m$ bonds, second factor describes the loss of translational entropy due to formation of $m$ bonds and the last factor is the number of ways to form $m$ bonds between $N$ particles and $N_s$ receptors. Substituting (\ref{z_langmuir}) into (\ref{free_langmuir}) and minimizing with respect to $m$ we get: 
\begin{equation}
    \frac{m}{N_s-m}=\frac{N-m}{V}ve^{-f}=cK
\end{equation}
Where $c$ is a bulk concentration of particles, and $K=ve^{-f}$ is equilibrium association constant between particle and receptor. It is clear that the expression obtained is exactly the Langmuir adsorption isotherm.

\begin{figure}
\includegraphics[width=0.7\columnwidth]{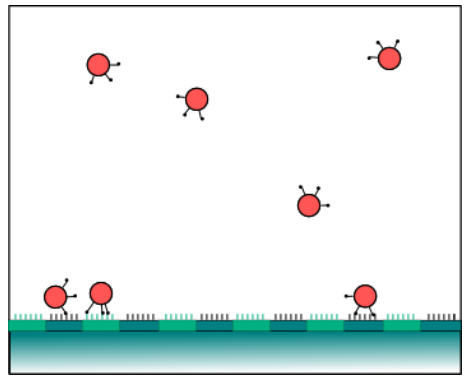}
\caption{Visual description of multivalent particles adsorption problem}
\label{multivalent}
\end{figure}
Now let us assume that each particle has $\kappa$ ligands, and the surface is divided into $N_s$ cells, each one of which has $n_R$ receptors (see Figure \ref{multivalent}) \footnote{We note here that we choose these model assumptions based on model already existing in the literature in order to establish connection. This approximation implies that the total coverage of the surface with particles is low, in other words adsorbed particles form dilute 2D-gas on the surface}. We also denote as $m_i$ the number of adsorbed particles which form $i$ bonds with the surface cell, so that the total number of adsorbed particles is $m=\sum_{i=1}^\kappa m_i$. The partition function due to bonds formation is therefore
\begin{eqnarray}
    Z_{ad}= &&e^{-f\sum im_i}\left(\frac{v}{V}\right)^{\sum m_i}\cdot \nonumber\\
    &&\cdot\frac{N!N_s!}{\left(N-\sum m_i\right)!\left(N_s-\sum m_i\right)!m_1!..m_\kappa!}\cdot \nonumber\\
    &&\cdot\prod_{i=1}^\kappa \left(\frac{\kappa!n_R!}{\left(\kappa-i\right)!\left(n_R-i\right)!i!}\right)^{m_i}
\end{eqnarray} 
The first factor accounts for energy of bonds, second for the loss of translational entropy of particles, next factor accounts for the number of ways to choose bonded particles and surface cells from the total number of particles and the final factor is a combinatorial number of ways to form $i$ bonds between ligands on the particle and receptors on the surface cell.  We substitute this expression into free energy and minimize it with respect to $m_i$ to get:
\begin{equation}
    \frac{m_i}{\left(N-m\right)\left(N_s-m\right)}=e^{-if}\frac{v}{V}\frac{\kappa!n_R!}{\left(\kappa-i\right)!\left(n_R-i\right)!i!}
\end{equation}
Next we make a standard assumption that $n_R>>\kappa$ \cite{Martinez-Veracoechea2011DesigningBinding}. We also denote $K_1=ve^{-f}$ association constant describing formation of the first bond and $K_c=e^{-f}$ association constant describing formation of subsequent bonds \cite{Curk2017DesignInteractions}. And summing all $m_i$ we get:
\begin{eqnarray}
    \frac{m}{\left(N_s-m\right)}=&&c\frac{K_1}{K_c}\sum_{i=1}^{\kappa}\frac{\kappa!n_R!}{\left(\kappa-i\right)!\left(n_R-i\right)!i!}K_c^i=\nonumber\\
    =&&c\frac{K_1}{K_c}\left(\left(1+K_c n_R\right)^\kappa-1\right)
    \label{final-MF}
\end{eqnarray}
which is the same expression as the main formula from the paper of Martinez-Veracoechea \textit{et al.} \cite{Martinez-Veracoechea2011DesigningBinding} assuming that $c\frac{K_1}{K_c}=cv=z$. Where $z$ is activity.
It is easy to calculate from it both surface coverage $\theta=\frac{m}{N_s}$ and superselectivity $\alpha=\frac{\partial \ln \theta}{\partial \ln n_R}$.

For linear polymers formation of each bond with receptor on the surface is accompanied by the loss of conformational entropy. However this entropy cost strongly depends on the length of linkers between stickers and the total length of the polymer. In the next section we will look into this.  

\section{Entropy cost of adsorption of a polymer chain}

In this paper we are interested in superselectivity which is associated with the fast increase of adsorbed material in the very beginning of adsorption process upon variation of the density of surface receptors. As such, we consider only the regime of adsorption of individual coils and do not take into account interactions between chains on the surface or the possibility of adsorption of several polymer coils on one surface patch \cite{deGennes1976ScalingAdsorption,Semenov1995StructureTails, Semenov1996AdsorptionSurface, Dubacheva2014SuperselectivePolymers, Tito2014OptimizingPolymers}.

In order to perform calculations I will use a Gaussian chain model. I intentionally concentrate on a Gaussian model of a polymer chain because it allows us to obtain an analytic result. We should note here that in general in order to correctly describe large-scale statistics of a polymer chain self-avoiding random walks should be used. From another point of view for superselective adsorption when stickers are not separated by very long polymer segments small-scale details of a polymer model, such as chain rigidity or excluded volume, are also important and are not grasped by Gaussian model as well. Despite of this Gaussian chain model proved to give qualitatively correct predictions in many cases and definitely helps in understanding of the effect of a conformational entropy on the behavior of polymer systems, which is our aim in the present study. 

Let us consider a Gaussian chain composed of $N$ statistical segments of size $a$ with ends located at points $\vec{r}$ and $\vec{r'}$.  The chain is described by the probability distribution function $p\left(\vec{r},\vec{r'};N\right)$ which is proportional to the number of chain conformations with ends constrained to positions $\vec{r}$ and $\vec{r'}$ \cite{Rubinstein2003PolymerPhysics, Matsen2006Self-ConsistentApplications}. This probability distribution function satisfies the diffusion equation, which can be intuitively understood from an analogy between an ideal polymer chain and the trajectory of Brownian particle:
\begin{equation}
    \frac{\partial p\left(r,r';N\right)}{\partial n}=\frac{a^2}{6}\Delta_r p\left(r,r';n\right)
\end{equation}
where $0<n<N$, $r$, $r'$ are not constrained, $p\left(r,r';0\right)=\delta\left(r-r'\right)$.
Solution of this equation is a Gaussian function:
\begin{equation}
    p\left(r,r';N\right)=\left(\frac{3}{2\pi Na^2}\right)^{3/2}e^{-\frac{3\left(r-r'\right)^2}{2Na^2}}
\end{equation}
With the probability distribution function satisfying normalization condition
\begin{equation}
    \int p\left(r,r';N\right) d^3r'=1
\end{equation}
If we are interested only in tracking one of the coordinates of chain ends, we integrate probability distribution function over other coordinates to obtain
\begin{equation}
    p\left(z,z';N\right)=\left(\frac{3}{2\pi Na^2}\right)^{1/2}e^{-\frac{3\left(z-z'\right)^2}{2Na^2}}
\end{equation}
Now, let's assume that there is a hard wall at $z_0=0$, then $p\left(z,z';N\right)$ satisfy the same equation
\begin{equation}
    \frac{\partial p\left(z,z';N\right)}{\partial n}=\frac{a^2}{6}\frac{\partial^2 p\left(z,z';N\right)}{\partial z^2} 
\end{equation}
but with boundary condition $p\left(z=0,z';N\right)=0$ (segments can't cross the surface). Solution is \cite{Wang2018ConformationalSurface} 
\begin{eqnarray}
    p\left(z,z';N\right)=&&
    \left(\frac{3}{2\pi Na^2}\right)^{1/2}\left(e^{-\frac{3\left(z-z'\right)^2}{2Na^2}}-\right.\nonumber\\
    &&\left.-e^{-\frac{3\left(z+z'\right)^2}{2Na^2}}\right);z,z'>0
\end{eqnarray}
If we integrate over location of the second end $z'$ \footnote{It is more correct to take integral between $a$ and $\infty$, but the difference is very small if $N$ is sufficiently large ($N>10$).}
\begin{equation}
    p\left(z;N\right)=\int_0^{\infty}p\left(z,z';N\right)dz'= erf \left(\sqrt{\frac{3}{2N}}\frac{z}{a}\right)
\end{equation}
we will get a "partition function" of a chain with one end free and another end fixed at position $z$ near the hard wall.
We can see that for the chain far away from the wall $p\left(z=\infty,N\right)=1$ as expected. 
\begin{figure}
    \includegraphics[width=0.7\columnwidth]{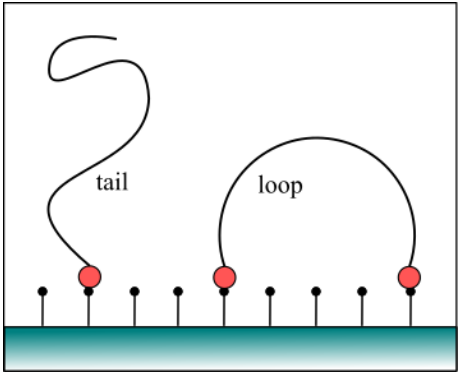}
    \caption{Schematic illustrating the "tail" and "loop" conformations}
    \label{tail-loop}
\end{figure}
\begin{figure}
    \includegraphics[width=0.7\columnwidth]{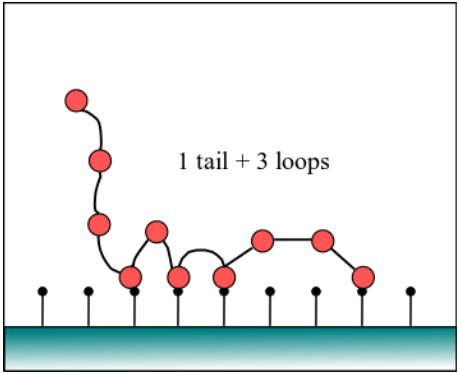}
    \caption{Schematic illustrating of the division of adsorbed polymer into tails and loops by adsorption segments}
    \label{tail-loop}
\end{figure}

We define a "tail" as a chain or a subchain that sticks with one end to the surface whilst another end is free (see Figure \ref{tail-loop}). The conformational free energy of a "tail" is 
\begin{eqnarray}
    \frac{F_{conf,tail}}{kT}=&&-\ln \left[erf \left(\sqrt{\frac{3}{2N}}\frac{z}{a}\right)\right]=\nonumber\\
    &&-\ln \left[erf \left(\sqrt{\frac{3}{2N}}\right)\right]
    \label{tail}
\end{eqnarray}
Here we assume that the coordinate of chain attachment point to the surface is $z=a$, the size of statistical segment.

If we call a "loop" a chain or a subchain with both ends fixed to the surface (see Figure \ref{tail-loop}) then the partition function of a loop is
\begin{eqnarray}
    p\left(z;N\right)=&&\int_0^{\infty}p\left(z,z';N\right)\delta \left(z-z'\right)dz'=\nonumber\\
    =&&\left(\frac{3}{2\pi N}\right)^{1/2}\left(1-e^{-\frac{6z^2}{Na^2}}\right)=\nonumber\\
    =&&\left(\frac{3}{2\pi N}\right)^{1/2}\left(1-e^{-6/N}\right)
\end{eqnarray}
in case $z=a$.

So, conformation free energy of "loop" is 
\begin{equation}
    \frac{F_{conf,loop}}{kT}=-\ln\left[\left(\frac{3}{2\pi N}\right)^{1/2}\left(1-e^{-6/N}\right)\right]
\end{equation}
 Now let us assume that adsorbed chain with length $N$ is split into two tails with lengths $N_1$ and $N_2$ and a loop with length $N_3$, $N=N_1+N_2+N_3$ by two adsorbed segments (see Figure \ref{polymer}). In this case the entropy cost of chain adsorption will be just a sum of contributions of tails and loops:
 \begin{eqnarray}
     F_{conf}=&&F_{conf,tail}\left(N_1\right)+F_{conf,loop}\left(N_3\right)+\nonumber\\
     +&&F_{conf,tail}\left(N_2\right)
 \end{eqnarray}

\begin{figure}
    \includegraphics[width=1\columnwidth]{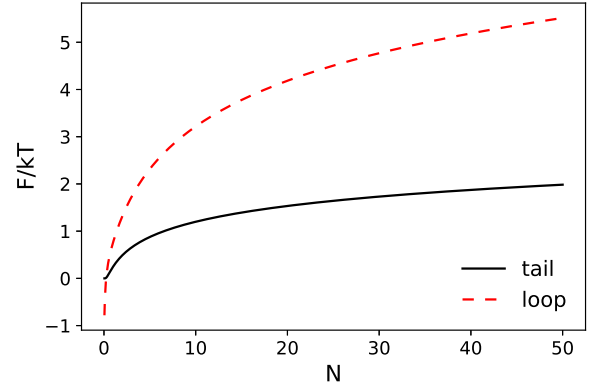}
    \caption{Dependence of free energy of tail (solid) and loop (dashed) on the number of segments they contain}
    \label{tail-loop-energy}
\end{figure}

It is useful to understand how loop and tail conformation free energies depend on the number of segments. Figure \ref{tail-loop-energy} shows that free energy of loop is always larger than for tail which is expected because tail due to one free end has larger number of conformations available compared to the loop.  We can also notice that tail free energy grows with $N$ much slower than linear. So, if we fix the number of segments bonded to the surface then the minimum of the total conformational free energy will correspond to the situation when loops have the minimum possible length and one tail long and the other short. 

\section{Selectivity of polymer adsorption}

We can see from the previous section that for the polymer chain not all ligands are equivalent (they can be distinguished by their location along the chain), so expression for $Z_{ad}$ needs to be modified. Let's assume that $\kappa$ is a number of stickers per polymer chain \footnote{We assume that only stickers interact with receptors on the surface. For other chain segments the wall is repulsive. Segments and stickers do not interact with each other.}. We assume that stickers are distributed uniformly and there are stickers at the ends of a chain (as in our simulation model). Let's denote as $\{\kappa_i\}$ some partition of $\kappa$ by $i$ adsorbed segments, characterized by lengths of tails and loops $\{n_1, n_2,...\}$.  The combinatorial number of ways to produce such partition is $\Omega\{\kappa_i\}$. We also denote as $m_{\{\kappa_i\}}$ the number of adsorbed chains, which are characterised by such partition. The total number of adsorbed chains is $m=\sum_{i=1}^{\kappa}\sum_{\{\kappa_i\}}m_{\{\kappa_i\}}$. We will also denote as $\prod\left[m_{\{\kappa_i\}}!\right]$ the product of factorials of all different numbers $m_{\{\kappa_i\}}$. Then for $Z_{ad}$ we have:
\begin{eqnarray}
    Z_{ad}=&&e^{-\sum_i i\epsilon\sum_{\{\kappa_i\}}m_{\{\kappa_i\}}}\left(\frac{R_g^{3}}{V}\right)^{m}\cdot\nonumber\\
    &&\cdot e^{-\sum_{i=1}^{\kappa}\sum_{\{n_i\}} m_{\{\kappa_i\}} F_{conf}\left(\{\kappa_i\}\right)}\cdot\nonumber\\
    &&\cdot\frac{N!N_s!}{\left(N-m\right)!\left(N_s-m\right)!\prod\left[m_{\{\kappa_i\}}!\right]}\cdot\nonumber\\ 
    &&\cdot\prod_{\{\kappa_i\}}\left(\Omega\left(\{\kappa_i\}\right)\frac{n_R!}{\left(n_R-i\right)!}\right)^{m_{\{\kappa_i\}}}
\end{eqnarray} 
Minimizing with respect to $m_{\{\kappa_i\}}$ we have 
\begin{equation}
    \frac{m_{\{\kappa_i\}}}{N_s-m}=cR_g^3e^{-i\epsilon}e^{-F_{conf}\left(\{\kappa_i\}\right)}n_R^i\Omega\left(\{\kappa_i\}\right)
\end{equation}
or summing all up we have:
\begin{equation}
    \frac{m}{N_s-m}=cR_g^3s\left(e^{-\epsilon}n_R\right)
    \label{m-final}
\end{equation}
Where 
\begin{equation}
    s\left(\gamma\right)=\sum_{i=1}^{\kappa}\gamma^i\sum_{\{\kappa_i\}}\Omega\left(\{\kappa_i\}\right)e^{-F_{conf}\left(\{\kappa_i\}\right)}
    \label{s-final}
\end{equation}
and $\gamma=n_Re^{-\epsilon}$. We can notice here that $s\left(t\right)$ is avidity for the linear polymer (compare with the formula 3.5 from Curk \textit{et al}.\cite{Curk2017DesignInteractions}). In the case when $F_{conf}=0$ this formula should converge to the formula for avidity of particles with no conformational degrees of freedom (we will henceforth refer to them as MV-particles) $s\left(\gamma\right)=\sum_{i=1}^{\kappa}\gamma^i\frac{\kappa!}{\left(\kappa-i\right)!i!}=\left(1+\gamma\right)^\kappa-1$. Let's demonstrate it for the example of a particle with valency $\kappa=3$, let's assume that two stickers are located at ends and one in the middle, length of linker is $n$. Then for the polymeric particle
\begin{eqnarray}
    s\left(n_Re^{-\epsilon}\right)=&&2n_Re^{-\epsilon-F_{tail}\left(2n\right)}+n_Re^{-\epsilon-2F_{tail}\left(n\right)}+\nonumber\\
    &&+2n_R^2e^{-2\epsilon-F_{tail}\left(n\right)-F_{loop}\left(n\right)}+\nonumber\\
    &&+n_R^2e^{-2\epsilon-F_{loop}\left(2n\right)}+\nonumber\\
    &&+n_R^3e^{-3\epsilon+2F_{loop}\left(n\right)}
\end{eqnarray}
It is clear that if we put all entropy terms to zero, this expression will converge to the one for tri-valent MV-particle
\begin{equation}
    s\left(n_Re^{-\epsilon}\right)=3n_Re^{-\epsilon}+3n_R^2e^{-2\epsilon}+n_R^3e^{-3\epsilon}
\end{equation}

Formulae (\ref{m-final}) and (\ref{s-final}) show the final result, in which sums are supposed to be taken numerically. Therefore the fraction of occupied surface sites is:
\begin{equation}
    \theta=\frac{m}{N_s}=\frac{cR_g^3s\left(e^{-\epsilon}n_R\right)}{1+cR_g^3s\left(e^{-\epsilon}n_R\right)}
\end{equation}
The selectivity with respect to the density of surface receptors then can be calculated as:
\begin{equation}
    \alpha=\frac{\partial \ln \theta}{\partial \ln n_R}=\frac{e^{-\epsilon}n_R\cdot s'\left(e^{-\epsilon}n_R\right)}{s\left(e^{-\epsilon}n_R\right)\left(1+cR_g^3 s\left(e^{-\epsilon}n_R\right)\right)}
\end{equation}

We can see that selectivity depends only on a cumulative parameter $\gamma=e^{-\epsilon}n_R$ and not on the density of receptors and the strength of interaction independently same as in the case of structure-less multivalent particles. 
So we will be using parameter $\gamma$ in all selectivity plots instead of $n_R$. Now we can analyse the obtained expressions for selectivity.

\begin{figure}
    \includegraphics[width=\columnwidth]{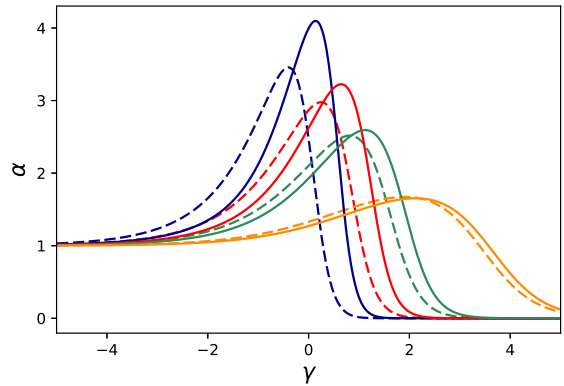}
    \caption{Comparison of selectivity for multivalent polymers (solid lines) with linker length $n=1$ with selectivity of Martinez-Veracoechae's multivalent particles (dashed lines).   Length of linkers n=1. Blue curves valency $\kappa=10$, red to $\kappa=6$, green to $\kappa=4$ and orange to $\kappa=2$. Concentration of the polymer solution is $cR_g^3=0.001$.}
    \label{sol-pol}
\end{figure}

\begin{figure}
    \includegraphics[width=\columnwidth]{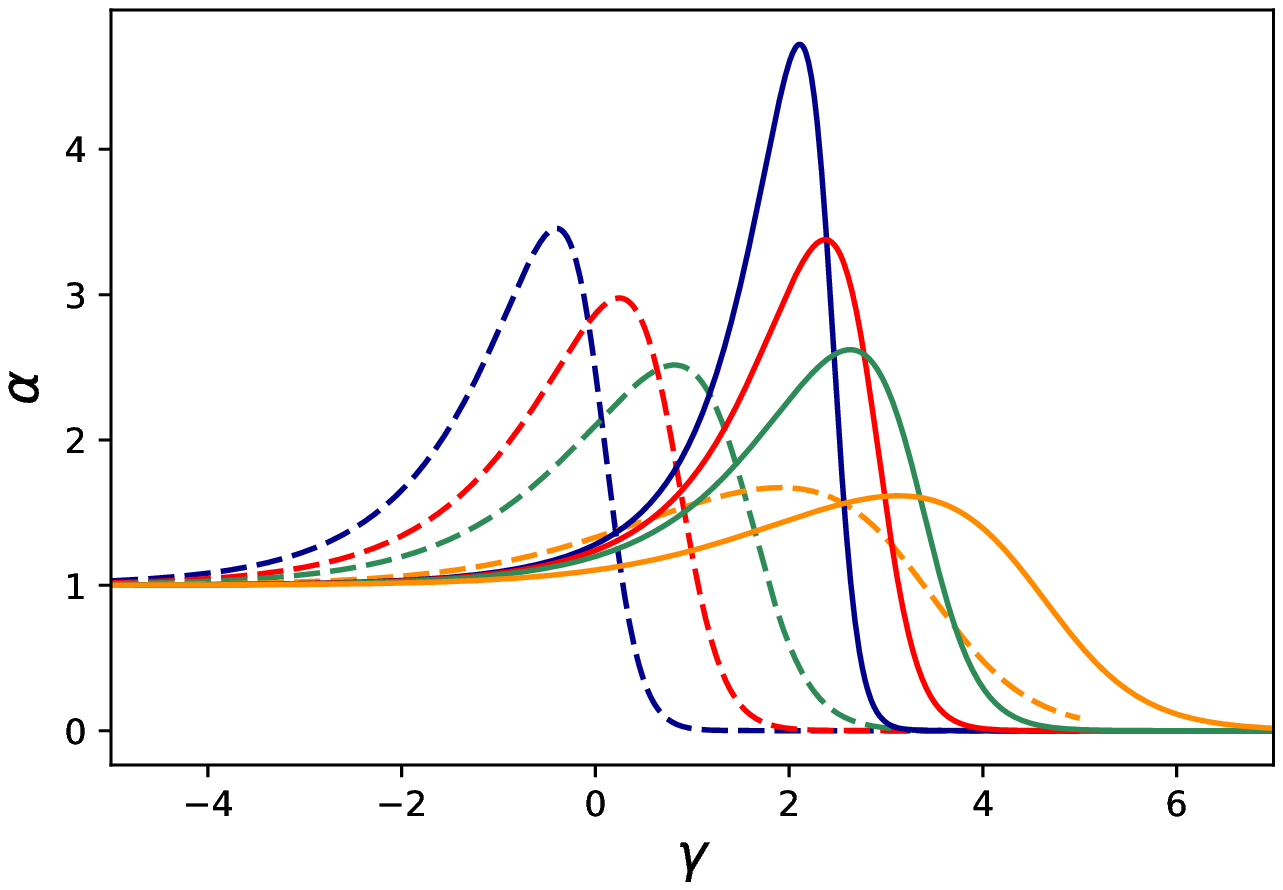}
    \caption{Comparison of selectivity for multivalent polymers (solid lines) with linker length $n=10$ with selectivity of Martinez-Veracoechae's multivalent particles (dashed lines).   Length of linkers n=1. Blue curves valency $\kappa=10$, red to $\kappa=6$, green to $\kappa=4$ and orange to $\kappa=2$.Concentration of the polymer solution is $cR_g^3=0.001$.}
    \label{sol-pol10}
\end{figure}

Figure \ref{sol-pol} shows dependence of selectivity on the valency of a polymer (solid curves). Calculations are done for the case when stickers are uniformly distributed along the chain, there are stickers at the chain ends and the length of linker is $N=1$ as in our simulation model (which will be discussed later). For comparison selectivity curves for VM-particles with the same valency are also plotted (dashed curves). First of all we can see that increase in valency leads to increase in selectivity. However selectivity for polymers is different from selectivity of MV-particles (Figure \ref{sol-pol}). For divalent particle the difference between the two is very small, however as the valency increases difference between selectivity of polymer and MV-particle increases. The main effect which we can observe is the shift of the selectivity peak towards higher values of $\gamma$ which can be explained by the fact that in order to pin a polymer at the surface, the strength of interaction should be larger compared to MV-particles to compensate for the loss of conformational entropy upon adsorption. If we increase the length of a linker (Figure \ref{sol-pol10}) we can see that peaks shift further to the right which corresponds to the increase in entropy of linkers.

For divalent particle there are only two stickers at the ends of a chain, and as we increase the length of a linker we increase a length of this chain. So, avidity for it is written as
\begin{equation}
    s\left(\gamma\right)=2\gamma e^{-F_{tail}\left(n\right)}+\gamma^2e^{-F_{loop}\left(n\right)}
\end{equation}
And we can re-write it as 
\begin{equation}
    s\left(\gamma'\right)=2\gamma'+{\gamma'}^2e^{-F_{loop}\left(n\right)+2F_{tail}\left(n\right)}
    \label{shift}
\end{equation}
So, we have new $\gamma'$ now which is responsible for the shift of the curve to the right, because $e^{-F_{tail}\left(n\right)}<1$, and change of the shape of the curve due to decrease of contribution of $\gamma'^2$ term due to the factor $e^{-F_{loop}\left(n\right)+2F_{tail}\left(n\right)}<1$. With increase in valency of a polymer the balance between different terms becomes more complicated. However, the shift is still predominantly defined by the smallest possible $F_{conf}$ which corresponds to the the attachment of the chain by one end. In this case $F_{conf}=F_{tail}\left(N\right)$ where $N$ is a length of the chain. All other factors affect the shape of a selectivity peak. If there are no stickers at the ends of the polymer, then the shift is determined by the sum of conformational energies of the longest and the shortest tails. 

\begin{figure}
    \includegraphics[width=\columnwidth]{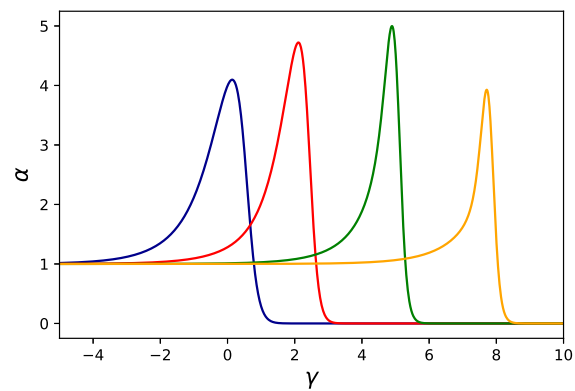}
    \caption{Selectivity curves for 10-valent polymers with linker lengths: blue curve $n=1$, red $n=10$, green $n=100$ and orange $n=1000$. Concentration of the polymer solution is $cR_g^3=0.001$.}
    \label{linker}
\end{figure}

From Figures \ref{sol-pol} and \ref{sol-pol10} we can see that for Gaussian polymer selectivity peak becomes sharper and larger in amplitude compared to the selectivity peak of MV-particles with the same valency. Additionally Figure \ref{linker} shows dependence of selectivity on the length of a linker for 10-valent polymer. We can see that as the linker length is increases the amplitude of the peak first increases before starting to decrease. However as discussed above this shape change strongly depends on the precise form of $F_{tail}$ and $F_{loop}$ and the balance between them which in turn is determined by a polymer model used. So this effect might not be present for other chain models or experimental systems.

\begin{figure}
    \includegraphics[width=\columnwidth]{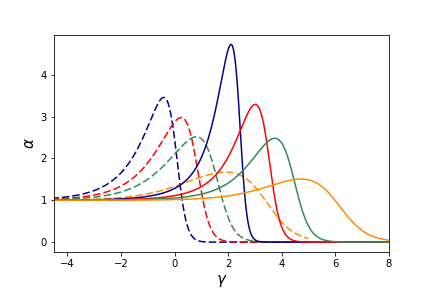}
    \caption{Comparison of selectivity for  multivalent polymers (solid line) and Martinez-Veracoechae's multivalent particles (dashed lines). Total length of polymer is fixed for all chains $N=90$, so length of linker depends on valency $n=N/\left(\kappa-1\right)$. Valencies are $\kappa=10$ for blue curve, $\kappa=6$ red, $\kappa=4$ green, $\kappa=2$ orange. Concentration of the polymer solution is $cR_g^3=0.001$.}
    \label{sol-pol-n-fix}
\end{figure}

Up to this point we considered polymeric particles with a fixed length of a linker but different valencies. This picture can be a realistic model in case when polymer is obtained by any supramolecular method by means of connecting oligomers into a polymer. In this case the total length of a polymer depends on its valency. A more common situation is when a polymer chain of a fixed length is modified by some number of ligands. It is therefore interesting to observe how selectivity changes when valency of a polymer of fixed length is changed, this case is considered in Figure \ref{sol-pol-n-fix}. Here we assume that the total length of the chain is $N=90$ and the length of linker is $n=N/\left(\kappa-1\right)$. We can see that in this case shift of the curve to the right is nearly independent of the polymer valency. This gives a support to the statement that the shift is mainly determined by free energy of the longest tail $F_{tail}\left(N\right)$.  


\begin{figure}
\includegraphics[width=\columnwidth]{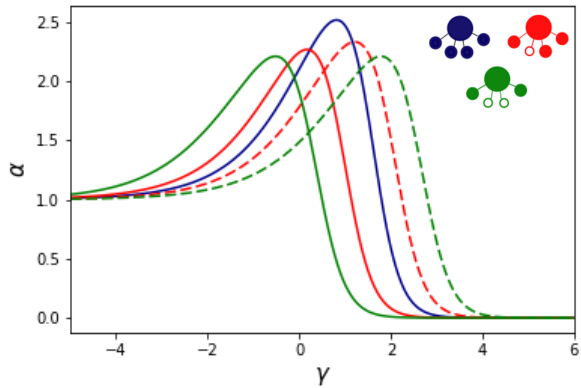}
\caption{The effect of modification of ligands on location of superselectivity peak for 4-valent MV-particles. Blue curve is unmodified 4-valent particle, red is modification of one ligand, green modification of two ligands. Dashed curves correspond to modification with weaker stickers, solid curves correspond to modification with stronger stickers. Concentration of the polymer solution is $cR_g^3=0.001$.}
\label{copolymer-MV}
\end{figure}

It is also interesting to look at particles with two types of ligands which have different binding energies to receptors on the surface. We assume that there is only one type of receptors. In order to simplify calculations we will consider 4-valent particles. Figure \ref{copolymer-MV} shows changes in selectivity curve for MV-particles when some fraction of ligands is modified. For new ligands association energy with receptors is $\epsilon_1=\epsilon+\ln\left(10\right)$ (solid red and green curves) or $\epsilon_1=\epsilon-\ln\left(10\right)$ (dashed red and green curves). We can see that modification of receptors leads to slight decrease in amplitude of selectivity peak and to its shift along $\gamma$ axis. Peak shifts to higher values of $\gamma$ for modification with weaker ligands and to lower values of $\gamma$ for modification with stronger ligands. The amplitude of shift increases with the number of ligands modified.

\begin{figure}
\includegraphics[width=\columnwidth]{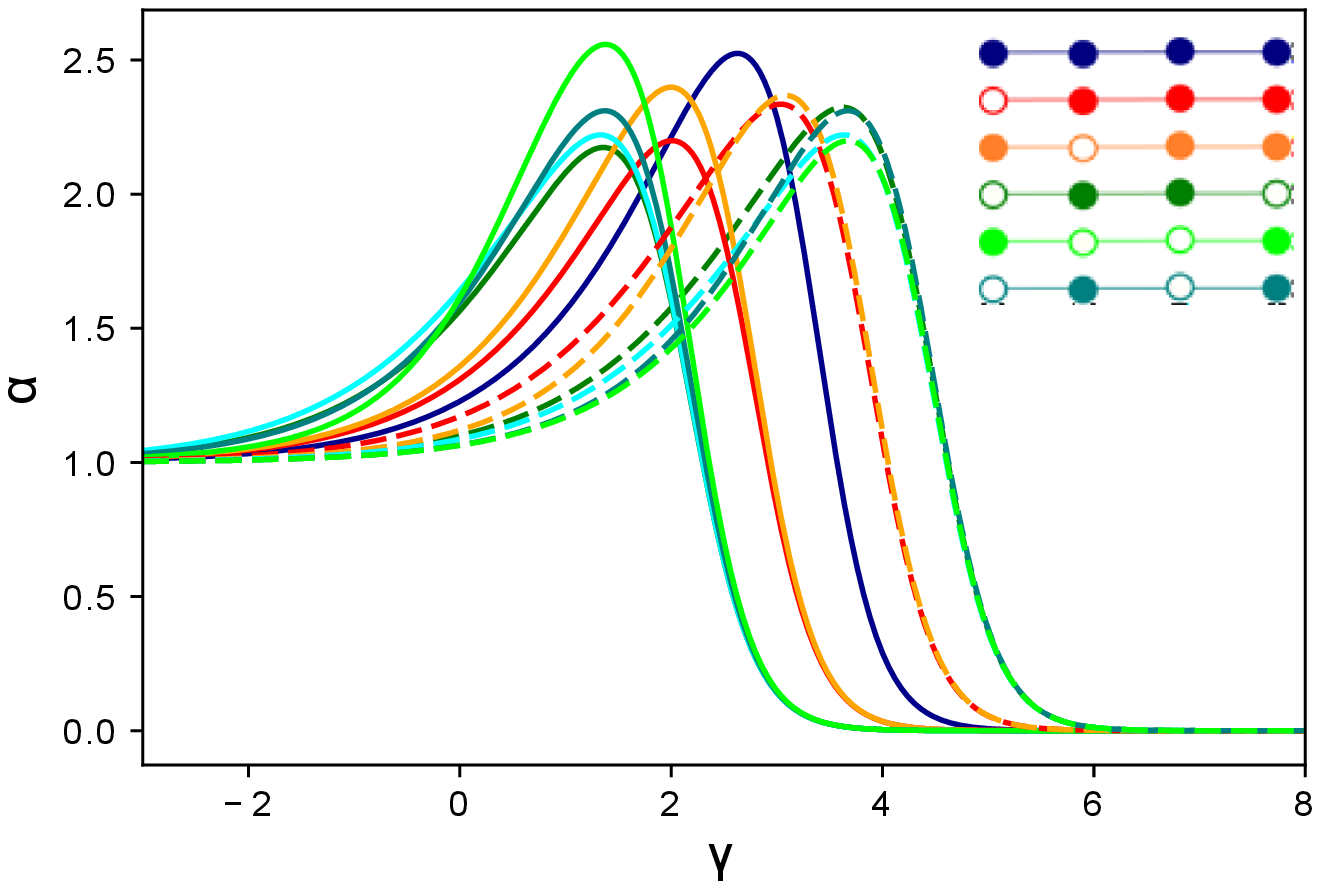}
\caption{The effect of modification of stickers on a polymeric 4-valent particle. Dark blue curve corresponds to unmodified polymer with valency $\kappa=4$ and linkers $n=10$. Other curves represent modified polymers. The colour represents modification pattern. Solid curves correspond to modification with stronger ligands and dashed curves correspond to modification with weaker ligands. Concentration of the polymer solution is $cR_g^3=0.001$.}
        \label{sequence-n10}
\end{figure}

\begin{figure}
    \includegraphics[width=\columnwidth]{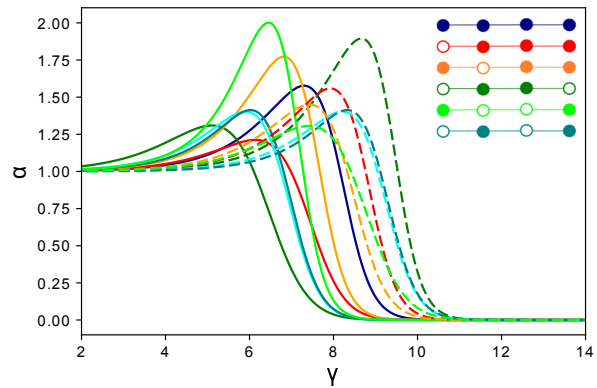}
    \caption{The effect of modification of stickers on a polymeric 4-valent particle with long linkers. Dark blue curve corresponds to unmodified polymer with valency $\kappa=4$ and linkers $n=1000$. Other curves represent modified polymers. The colour represents modification pattern. Solid curves correspond to modification with stronger ligands and dashed curves correspond to modification with weaker ligands. Concentration of the polymer solution is $cR_g^3=0.001$.}
    \label{sequence-n1000}
\end{figure}

In a polymeric particle different ligands are not equivalent, so in general modification pattern is determined not only by the number of ligands modified but also by their sequence. Changes in selectivity curve for 4-valent polymeric particle with length of linker $n=10$ are shown in Figure \ref{sequence-n10}. We can see as in the previous case peak shifts to the left along $\gamma$ axis in case of modification with stronger stickers and to the right in opposite situation. We can also see that surprisingly sequence is not important and location of the peak is mainly determined by the number of ligands modified. However it is true only up to a certain point. 
Figure \ref{sequence-n1000} shows the effect of sequence for a polymer with linkers with length $n=1000$. We can see that in this case selectivity curves for polymers with the same sticker profile but different sequences are different. 

\section{Conclusions}

In this paper we developed an analytical approach to predict selectivity of adsorption of polymeric multivalent particles. As a byproduct of these calculations avidities are also obtained. One of the main advantages of the proposed approach is that it allows rapid estimation of avidity and selectivity for polymeric particles, which can be applied to polymers of any architecture, without the need to set up a large scale simulation and get an intuitive understanding of the main parameters directing the behavior of the system.

We applied developed theory to show that though ligands on a linear polymer are not equivalent, superselectivity is present for multivalent polymers. Selectivity increases with increase in valency of a polymer. Adsorption of multivalent polymers starts at higher values of $\gamma$ compared to structureless particles with the same valency because additional attraction strength or density of receptors is needed to compensate the loss of conformational entropy upon adsorption. The shift of the onset of adsorption can be roughly estimated as $\gamma'=\gamma\cdot e^{-F_{tail}\left(N\right)}$ where $\gamma$ corresponds to the onset of adsorption of particles with the same valency but no internal degrees of freedom and $F_{tail}\left(N\right)$ is the conformational energy of a polymer attached by one end to a surface. 

We also showed that if a multivalent polymer has ligands of two types and linkers between ligands are not too long then superselectivity curve depends only on the number of different ligands and does not depend on their sequence. Modification of the polymer with weaker ligands shifts superselectivity peak to larger $\gamma$ and with stronger ligands to smaller values of $\gamma$. This is a particularly interesting finding  which shows that straightforward modification of polymer with given ligands without control of their sequence should be a working strategy allowing to produce particles targeting required surfaces.  

We also established connection between the field of multivalent particles and statistical theory of associating fluids. Methods developed for associating polymers\cite{Semenov1998ThermoreversibleStatics} and fluids can be successfully used for considering more complicated systems of particles, for example when self-association is present and it competes with cross-association\cite{Coleman1995}. Also there are ways to account for cooperativity, when interaction energy depends on the number of bonds formed\cite{Veytsmant1993, Patyukova2018HydrogenExperiment}.

\begin{acknowledgments}
EP is thankful to Mark A. Miller, Halim Kusumaatmaja and Andrew T.R. Christy for sharing the knowledge about multivalecy and superselectivity and discussions of this work, and to Martin Greenall  for reading the manuscript.  
\end{acknowledgments}

\bibliography{Mendeley}

\end{document}